\documentclass[twocolumn,showpacs]{revtex4}

\usepackage{amsmath,epsfig,multirow,delarray}
\begin{document}

  \newcommand{\ftimes}{}          

\preprint{CGPG-03/8-1}

\title{Numerical stability of the AA evolution system \\ compared to
the ADM and BSSN systems}

\author{Nina Jansen, Bernd Br\"ugmann, Wolfgang Tichy}
\affiliation{
Center for Gravitational Physics and Geometry and 
Center for Gravitational Wave Physics\\
Penn State University, University Park, PA 16802
}

\date{October 20, 2003}

\begin{abstract}
We explore the numerical stability properties of an evolution system
suggested by Alekseenko and Arnold. We examine its behavior on a set
of standardized testbeds, and we evolve a single black hole with
different gauges. Based on a comparison with two other evolution
systems with well-known properties, we discuss some of the strengths and
limitations of such simple tests in predicting numerical stability
in general.
\end{abstract}

\pacs{
04.25.Dm   
%
\quad Preprint number: CGPG-03/8-1
}

\maketitle

\section{Introduction}

In recent years the quest for finding a numerically stable formulation of
the Einstein evolution equations has become more and more intense, see
e.g.~\cite{Reula98a,Friedrich:2000qv} for reviews. Effort has been put into
finding first order symmetric hyperbolic formulations of the evolution
equations, since the properties of such systems can be analyzed
mathematically. However, even if an evolution system is symmetric hyperbolic
there is no guarantee that its numerical implementation is stable when
evolving a highly dynamic black hole spacetime. The
Baumgarte-Shapiro-Shibata-Nakamura (BSSN)~\cite{Shibata95,Baumgarte99}
system is an example of a system that is not first order symmetric
hyperbolic but has nice stability properties. Thus, current mathematical
analysis is not sufficient to explore the properties of an evolution system.
The system must be implemented and tested numerically before we can draw
definite conclusions about its viability for a particular physical
application.

In \cite{Alekseenko2002}, Alekseenko and Arnold (AA) suggest a first order
formulation of the evolution equations that is symmetric hyperbolic when
considering only a subset of the variables. In particular, the metric itself
and some of its first spatial derivatives are not considered part of the
evolution system and are treated as given functions when showing
hyperbolicity. It is argued that this is sensible since the metric is
derivable from an ordinary differential equation. A distinguishing feature
of the AA system is that a minimal number of first derivatives of the metric
are introduced as independent variables. The system has only 20 unknowns and
no parameters that have to be fixed for hyperbolicity, so it is relatively
simple for a symmetric hyperbolic system.

We have chosen to implement the AA system numerically and to compare
its numerical properties with those of the Arnowitt-Deser-Misner (ADM)
\cite{Arnowitt62,York79} and BSSN systems. The ADM system is known to be
unstable in many situations, but we include it here since for finite
time intervals of evolutions it has been used rather successfully in
practice, e.g.\ in 3D black hole
simulations~\cite{Anninos94c,Bruegmann97,Brandt00,Alcubierre00b,Baker00b,Baker:2001nu,Baker:2002qf,Baker:2003ds}. The
BSSN system can be obtained from a trace-conformal decomposition of
the ADM equations, and with suitable techniques it is very stable for
black hole evolutions. For example, the first stable evolution for
more than $100000M$ of a single Schwarzschild black hole in a
(3+1)-dimensional code without adapted coordinates was obtained with a
modified BSSN system~\cite{Alcubierre00a}.
Both the ADM and BSSN systems are not first order. However, there exist first
order versions of the BSSN system which are symmetric 
hyperbolic~\cite{Alcubierre99c,Frittelli99,Friedrich:2000qv,Sarbach02a}, if the
densitized lapse and shift are considered given functions.
Straightforward first order forms of ADM are only
weakly hyperbolic, which implies certain numerical instabilities
(see e.g.~\cite{Calabrese02a}). 

The second order version of the BSSN system shares with the AA system
the property that a subsystem of it is indeed symmetric hyperbolic. A
crucial step in the construction of BSSN is the introduction of the
contracted Christoffel symbol of the conformal metric,
$\tilde\Gamma^i$, as an independent variable. The evolution equation
for the extrinsic curvature then contains derivatives of the metric
only in the form of a Laplace operator, so that the metric obeys a
wave equation if $\tilde\Gamma^i$ is considered a prescribed variable,
ignoring its own evolution equation.  This partial hyperbolicity of
the BSSN system may well be a crucial ingredient in its success as
evolution system.  Hence the question arises whether symmetric
hyperbolicity in a subsystem implies a numerical advantage for the AA
system.

There is a variety of numerical tests that one can perform on an
evolution system. A complex and important issue is that seemingly
minor changes in the test conditions and implementation of the system
can lead to very different conclusions about
stability. In~\cite{Alcubierre2003:mexico-I}, an important step is
taken towards creating a set of tests that can serve as a standardized
benchmark for stability. Our discussion of the AA system in comparison
to ADM and BSSN can also be viewed as a contribution to the ongoing
development of such benchmarks. On the one hand, we report on the
performance of our particular implementation of ADM and BSSN, where a
large body of prior work allows us to judge how representative the
current benchmark is. On the other hand, we apply these tests to a
new evolution system about which nothing is known so far from numerical
experiments, and the question is, for example, whether one can predict the
usefulness of the AA system for black hole evolutions. We therefore
also discuss results for the evolution of a single black hole that
go beyond the current set of benchmarks in~\cite{Alcubierre2003:mexico-I}. 

The paper is organized as follows. In Sec.~\ref{secAA}, we write out
the AA system explicitly, in notation that is more familiar to
numerical relativists, and we show that the complete system is not
symmetric hyperbolic. We introduce a time-independent conformal
rescaling of the AA system that we will need later for black hole
evolutions. In Sec.~\ref{secTests}, we perform three tests from the
recently proposed test suite \cite{Alcubierre2003:mexico-I} for
numerical evolutions, the robust stability test, the gauge stability
test, and the linear wave stability test. Finally, we evolve a single
black hole space time in Sec.~\ref{secBH}, since this is the physics
example that we are most interested in. We conclude with a discussion
in Sec.~\ref{secDiscuss}.

\section{The AA system}
\label{secAA}

In~\cite{Alekseenko2002}, the evolution system is given in compact
notation. Here we write it out in the form in which we have
implemented it:
\begin{widetext}
\begin{eqnarray}
\label{AAeqn_first}
f_{ijk} &=& \frac{1}{2\sqrt{2}} \left(g_{ik,j} - g_{jk,i} + \left(\left(g_{pj,q}-g_{pq,j}\right)g_{ik} - \left(g_{pi,q} - g_{pq,i}\right)g_{jk}\right) g^{pq} \right) ,\\
w_{ij} &=& \frac{1}{2}\left(\beta^{p}_{\;,j} g_{pi} + \beta^{p}_{\;,i}
g_{pj}\right) ,\\
K &=& g^{ij} K_{ij} ,\\
c^{1)}_{ij} &=& \frac{1}{4} \left(2 g_{pq,j} g^{pq}_{\;\;,i} - g_{pq,i} g^{pq}_{\;\;,j} + 2 g_{pj,i} g^{pq}_{\;\;,q} - 2 g_{pj,q} 
\left(g^{pq}_{\;\;,i} + g_{ri,s} \left(g^{ps} g^{qr} - g^{pr}
g^{qs}\right)\right) - \right.\\ \nonumber
&\quad& \left. g_{ij,s} g_{pq,r} g^{pq} g^{rs} + g_{pq,r} g_{si,j}
g^{pq} g^{rs} + g_{pq,r} g_{sj,i} g^{pq} g^{rs}\right) ,\\
c^{2)}_{ij} &=& \frac{1}{4}\left(2 c^{1)}_{ij} + 2 c^{1)}_{ji} - g_{pj,q} g^{rs}_{\;\;,s} g_{ri} g^{pq} + 
g_{pq,j} g^{rs}_{\;\;,s} g_{ri} g^{pq} - g_{pi,q} g^{rs}_{\;\;,s} 
g_{rj} g^{pq} + g_{pq,i} g^{rs}_{\;\;,s} g_{rj} g^{pq} + \right. \\
\nonumber &\quad& \left. 
g_{pq,r} g^{rs}_{\;\;,j} g_{si} g^{pq} + g_{pq,r} g^{rs}_{\;\;,i} 
g_{sj} g^{pq} - g_{pq,r} g^{qs}_{\;\;,j} g_{si} g^{pr} - 
g_{pq,r} g^{qs}_{\;\;,i} g_{sj} g^{pr} - g_{pj,q} g^{qr}_{\;\;,s} 
g_{ri} g^{ps} - \right. \\
\nonumber &\quad& \left. 
g_{pq,j} g^{qr}_{\;\;,s} g_{ri} g^{ps} - 
g_{pi,q} g^{qr}_{\;\;,s} g_{rj} g^{ps} - g_{pq,i} g^{qr}_{\;\;,s} 
g_{rj} g^{ps} + 2 g_{pj,q} g^{pr}_{\;\;,s} g_{ri} g^{qs} + 
2 g_{pi,q} g^{pr}_{\;\;,s} g_{rj} g^{qs} - \right. \\
\nonumber &\quad& \left. 
4 g_{ij,s} g_{pq,r} g^{pr} g^{qs} + 
4 g_{ij,s} g_{pq,r} g^{pq} g^{rs}\right) ,\\
c^{4)}_{ij} &=& \frac{1}{2}\left(2\alpha c^{2)}_{ij} - 2 \alpha_{,i,j} - \alpha_{,p} g_{ij,q} g^{pq} + 
\alpha_{,p} g_{qi,j} g^{pq} + \alpha_{,p} g_{qj,i} g^{pq} + 
2\alpha K K_{ij} + \right. \\
\nonumber &\quad& \left.
2 \beta^{p}_{\;,j} K_{pi} + 2 \beta^{p}_{\;,i} K_{pj} - 
4\alpha g^{pq} K_{pi} K_{qj}\right) ,\\
c^{5)} &=& -\frac{1}{4} \left(g_{pq,r} \left(2 g^{qr}_{\;\;,s} g^{ps} - g^{pq}_{\;\;,s} g^{rs} + g_{st,u} g^{pq} g^{ru} g^{st}\right)\right) \\
B_{ij} &=& \frac{1}{2} \left(2 c^{4)}_{ij} -\alpha \left(g_{ij,s} g_{pq,r} \left(-\left(g^{ps} g^{qr}\right) + 
g^{pq} g^{rs}\right) + g_{ij} \left(c^{5)} + K^2 - K_{pq}
K^{pq}\right)\right)\right) , \\
c^{6)}_{ijk} &=& \frac{1}{2 \sqrt{2}} \left(\beta^{p}_{\;,q} \left(g_{rj,p} g_{ik} - g_{ri,p} g_{jk}\right) g^{qr} + 
\beta^{p}_{\;,j} \left(g_{ik,p} - g_{qr,p} g_{ik}
g^{qr}\right) + \beta^{p}_{\;,i} \left(-g_{jk,p} + g_{qr,p} g_{jk}
g^{qr}\right)\right) ,\\
c^{7)}_{ijk} &=& \frac{1}{2} 
\left(2 c^{6)}_{ijk} + \sqrt{2} \left(w_{ik,j} - w_{jk,i} + 
w_{pj,q} g_{ik} g^{pq} - w_{pq,j} g_{ik} g^{pq} - 
w_{pi,q} g_{jk} g^{pq} + w_{pq,i} g_{jk} g^{pq} -  \right. \right. \\
\nonumber &\quad& \left. \left.
\alpha \left( g_{pj,q} g^{pq} K_{ik} + g_{pq,j} g^{pq} K_{ik} + 
g_{pi,q} g^{pq} K_{jk} - g_{pq,i} g^{pq} K_{jk} + 
g_{pj,q} g_{ik} K^{pq} - g_{pq,j} g_{ik} K^{pq} - \right. \right. \right.\\
\nonumber &\quad& \left. \left. \left.
g_{pi,q} g_{jk} K^{pq} + g_{pq,i} g_{jk} K^{pq}\right) + 
g_{pj,q} g^{pq} w_{ik} - g_{pq,j} g^{pq} w_{ik} - 
g_{pi,q} g^{pq} w_{jk} + g_{pq,i} g^{pq} w_{jk} - \right. \right.\\
\nonumber &\quad& \left. \left. 
g_{pj,q} g_{ik} w^{pq} + g_{pq,j} g_{ik} w^{pq} + 
g_{pi,q} g_{jk} w^{pq} - g_{pq,i} g_{jk} w^{pq}\right)\right) ,\\
C_{ijk} &=& \frac{1}{4} \left(4 c^{7)}_{ijk} + \sqrt{2} \left(2 K \alpha_{,j} g_{ik} - 2 K \alpha_{,i} g_{jk} - 
2\alpha g^{pq}_{\;\;,q} g_{jk} K_{pi} + 2\alpha g^{pq}_{\;\;,q} g_{ik} 
K_{pj} -\alpha g^{pq}_{\;\;,j} g_{ik} K_{pq} + \right. \right.\\
\nonumber &\quad& \left. \left.
\alpha g^{pq}_{\;\;,i} g_{jk} K_{pq} + 2 \alpha_{,p} g_{jk} g^{pq} 
K_{qi} - 2 \alpha_{,p} g_{ik} g^{pq} K_{qj} - 
\alpha g_{pq,r} g_{jk} g^{pq} g^{rs} K_{si} + 
\alpha g_{pq,r} g_{ik} g^{pq} g^{rs} K_{sj}\right)\right) ,\\
\partial_0 g_{ij} &=& - 2 \alpha K_{ij} + 2 w_{ij} ,\\
\partial_0 K_{ij} &=& \frac{1}{2} \left(2 B_{ij} + \sqrt{2}\alpha \left(g^{pq}_{\;\;,q} \left(f_{pij} + f_{pji}\right) + 
f_{pij,q} g^{pq} + f_{pji,q} g^{pq} - 
g^{pq}_{\;\;,r} f_{psj} g_{qi} g^{rs} + g^{pq}_{\;\;,r} f_{sjp} 
g_{qi} g^{rs} - \right. \right.\\
\nonumber &\quad& \left. \left.
g^{pq}_{\;\;,r} f_{psi} g_{qj} g^{rs} + 
g^{pq}_{\;\;,r} f_{sip} g_{qj} g^{rs}\right)\right) , 
\\
\partial_0 f_{ijk} &=& C_{ijk} + \frac{1}{\sqrt{2}} \left(-\alpha
K_{ik,j} +\alpha K_{jk,i} - \alpha_{,j} K_{ik} + \alpha_{,i}
K_{jk}\right) .
\label{AAeqn_last}
\end{eqnarray}
\end{widetext}
The operator $\partial_0 = \partial_t - \beta^p\partial_p$ is defined
in terms of ordinary partial derivatives for all variables.

The new variable $f_{ijk}$ is anti-symmetric in $i$ and $j$ and also
satisfies the cyclic property $f_{ijk} + f_{jki} + f_{kij} = 0$. This
means that $f_{ijk}$ represents 8 independent variables. The total
number of 20 variables is thus smaller than in the case of most first
order formulations (which usually involve 30 or more variables), but
larger than for the BSSN system, which has 15 independent variables.

In addition, the equations in the AA system are more complex than in
the BSSN system, so that the AA evolution system takes roughly twice
as long to run the same number of iterations, even though we have made
an effort to implement the AA system in a numerically optimal way. The
simulations were carried out with the BAM code, which is a rewritten
version of the code used in~\cite{Bruegmann97}.  Part of
BAM is a Mathematica script to convert tensor equations to C code, and
the AA equations were generated this way using Mathematica and
MathTensor.

\subsection{Hyperbolicity of the AA system}
\label{sec:hyper}
Looking at Eqs.\ (\ref{AAeqn_first})-(\ref{AAeqn_last}) 
and assuming that $\alpha$ and $\beta^i$ are prescribed given functions, 
we see that the system has the form
\begin{equation}
\label{AA_form}
\partial_t u + A^i(u) u_{,i} + v(u)
+ w\left(m^{ij}(u) u_{,i}^T Q(u) u_{,j} \right) = 0 .
\end{equation}
Here 
\begin{equation}
u=\begin{pmatrix} 	g \\ 
			K \\ 
			f \\ \end{pmatrix}
\end{equation}
is the solution vector with spatial indices suppressed, and
\begin{equation}
A^i(u)=\begin{pmatrix}	0	& 0	& 0 \\
			r^i(u)	&a^i(u)	& c^i(u) \\
			s^i(u)  &c^i(u) & b^i(u) \\ \end{pmatrix}
\end{equation}
is a matrix which contains a symmetric submatrix 
\begin{equation}
S^i(u)=\begin{pmatrix}	a^i(u)	& c^i(u) \\
			c^i(u)	& b^i(u) \\ \end{pmatrix} .
\end{equation}
In addition, Eq.~(\ref{AA_form}) contains the two vector valued 
functions $v$ and $w$, where the argument of $w$ depends on
\begin{equation}
u_{,i}=\begin{pmatrix}	g_{,i} \\
        		K_{,i} \\
        		f_{,i} \\ \end{pmatrix}
\end{equation}
and its transpose $u_{,i}^T$,
and also on another matrix $Q(u)$, 
which is of the simple form
\begin{eqnarray}
Q(u)=\begin{pmatrix}	q(u) 	& 0 	& 0 \\
			0 	& 0	& 0 \\ 
			0	& 0	& 0 \\ \end{pmatrix} ,
\end{eqnarray}
so that the argument of $w$ in fact only depends on squares of first
derivatives of $g_{ij}$. 

From Eq.~(\ref{AA_form}) it is immediately apparent that the system is
of first order form, as no second derivatives of $u$ appear.
Nevertheless the system differs from the standard first order form by
the term $w$. This means that the standard theorems about
well-posedness (see e.g.~\cite{Reula98a})
are not applicable, and that we cannot easily compute
characteristic speeds and modes of the system.  However, if we
linearize the system around any background $u^B$, all terms of the
form $w$ drop out. This can be seen as follows. Assume that
\begin{equation}	 
\label{linearAnsatz}     
u=u^B + u^L = \begin{pmatrix}	g^B +g^L \\
                       		K^B +K^L \\
                       		f^B +f^L \\ \end{pmatrix} , 
\end{equation}           
where $u^L$ is a small perturbation to the background $u^B$.
Then 
\begin{eqnarray}
&& w\left(m^{ij}(u) u_{,i}^T Q(u) u_{,j} \right) \nonumber \\
&&= w\left(m^{ij}(u) g_{,i} q(u) g_{,j} \right) \nonumber \\
&&= w\left(m^{ij}(u) g^B_{,i} q(u) g^B_{,j} \right) + \nonumber \\
&&\quad\;\; w\left(m^{ij}(u) ( g^B_{,i} q(u) g^L_{,j} + 
         g^L_{,i}q(u) g^B_{,j}) \right) + O\left( (g^L)^2 \right) \nonumber \\
&&= w\left(m^{ij}(u) g^B_{,i} q(u) g^B_{,j} \right) + 
    d^i(u^B) g^L_{,i} +O\left( (g^L)^2 \right),
\end{eqnarray}  	
and hence Eq.~(\ref{AA_form}) becomes
\begin{equation}
\label{AA_form_linear}
\partial_t u^L + \bar{A}^i u^L_{,i} + \bar{v}(u^L) 
+ O\left( (u^L)^2 \right) + \bar{w}(u^B) = 0 ,
\end{equation}
which is now of standard first order form, but with a modified 
matrix
\begin{equation}
\label{A_linear}
\bar{A}^i=\begin{pmatrix}	0	  & 0	    & 0 \\
				\bar{r}^i &a^i(u^B) & c^i(u^B) \\
				\bar{s}^i &c^i(u^B) & b^i(u^B) \\ \end{pmatrix}
.
\end{equation}
Looking at Eq.~(\ref{AA_form_linear}) and (\ref{A_linear}), it is
immediately clear that the system is not symmetric hyperbolic.  In
fact, since the matrix $S^i(u^B)$ has several zero eigenvalues,
the full system in general is unlikely to have a complete set of
eigenvectors and thus to be strongly hyperbolic.

The exception is linearizion around flat space where 
$\bar{r}^i$ and $\bar{s}^i$ 
in Eq.~(\ref{A_linear}) vanish, so that the system
(\ref{AA_form_linear}) becomes symmetric hyperbolic in this special
case.  Also, since $S^i(u)$ is symmetric, the subsystem of $K_{ij}$
and $f_{ijk}$ with $g_{ij}$ considered as a prescribed variable is
symmetric hyperbolic, which holds true even for 
the non-linearized system (\ref{AA_form}). 
Alekseenko and Arnold \cite{Alekseenko2002} emphasize this last point and
stress that the evolution equation for $g_{ij}$ is (as usual) just an
ordinary differential equation. Yet in numerical evolutions the latter
property may be unimportant, since
$g_{ij}$ has to be evolved along with the other variables and cannot be
prescribed. In addition, it is not clear if the fact that the evolution
system (\ref{AA_form}) contains a symmetric hyperbolic subsystem has any
bearing on the stability properties of the full system.

\subsection{Conformal version of the AA system}

In Sec.~\ref{secBH}, we report on numerical evolutions of black hole
puncture data \cite{Brandt97b,Bruegmann97,Alcubierre02a}.
In order to numerically evolve puncture data without excision the AA
system has to be modified.  The reason is that e.g.\ for two punctures
the metric has the form
\begin{equation}
\label{punc_metric}
g_{ij} = \phi^{4} \bar{g}_{ij}
\end{equation}
with conformal factor 
\begin{equation}
\phi = 1 + \frac{m_1}{2r_1} + \frac{m_2}{2r_2} + u ,
\end{equation}
where $r_1$ and $r_2$ are distances from the punctures, $m_1$ and
$m_2$ are the bare puncture masses, and $u$ is finite.  From
Eq.~(\ref{punc_metric}) it is apparent that the physical metric
diverges like
\begin{equation}
g_{ij} \sim \frac{1}{r^4} 
\end{equation}
at each puncture, so that finite differencing calculations across or
near any puncture will fail. The same problem also occurs in the
variable
\begin{equation}
f_{ijk} \sim (g_{ik,j} - g_{jk,i}) \sim \frac{1}{r^5} .
\end{equation}
For this reason we do not evolve the variables $g_{ij}$ and $f_{ijk}$
directly, rather we rescale $g_{ij}$ and $f_{ijk}$ by the time
independent conformal factor
\begin{equation}
\psi = 1 + \frac{m_1}{2r_1} + \frac{m_2}{2r_2} 
\end{equation}
such that the rescaled quantities 
\begin{eqnarray}
\label{scaled_g}
\bar{g}_{ij}  &=& \psi^{-4} g_{ij}, \\
\label{scaled_f}
\bar{f}_{ijk} &=& \psi^{-6} f_{ijk}
\end{eqnarray}
become finite at the puncture. 
Furthermore, we also introduce the rescaled extrinsic curvature
\begin{equation} 
\label{scaled_K}	
\bar{K}_{ij} = \psi^{-4} K_{ij} ,
\end{equation}
in order to remove divergences in $K_{ij}$, which in the case of
punctures with spin behaves as
\begin{equation}
{K}_{ij} \sim \frac{1}{r} .
\end{equation}

Since $\psi$ is an a priori prescribed time independent function,
the principal part of the system of evolution equations
(\ref{AAeqn_first})-(\ref{AAeqn_last}) remains unchanged if we use the
rescalings in Eqs.~(\ref{scaled_g}), (\ref{scaled_f}), and
(\ref{scaled_K}) to express the system in terms of the new variables
$\bar{g}_{ij}$, $\bar{K}_{ij}$, and $\bar{f}_{ijk}$.  In addition, all
terms involving $\bar{g}_{ij}$, $\bar{K}_{ij}$, $\bar{f}_{ijk}$, and
their derivatives are finite in the new system so that finite
differencing can be used without trouble.  There are, however,
additional terms containing spatial derivatives of $\psi$, which
cannot be computed using finite differencing. Yet, since $\psi$ is a
known function we can use analytic expressions for its
derivatives. Furthermore, all such derivative terms also contain
appropriate powers of $\psi$ which make them finite.

\section{Stability tests}
\label{secTests}

We have used some of the stability testbeds suggested in
\cite{Alcubierre2003:mexico-I}, also known as the ``apples with apples'' tests,
to explore the properties of the AA system. We also show test results
for the ADM and BSSN systems for comparison.

\subsection{Robust stability test}

The purpose of the robust stability test is to determine how an
evolution system will handle random errors that are bound to occur at
machine-precision level. Random constraint violating initial data in
the linearized regime is used to simulate this machine error.

All apples with apples test are run on a full-3D grid, but in this
case with periodic boundary conditions in all directions. The 3D
domain is a 3-torus, and here we only use 3 grid-points in the $y$- and
$z$-directions. The parameters are:
\begin{itemize}
\item Simulation domain: $x \in [-0.5, +0.5]$
\item Number of grid points in each direction: 
  $n_x=50\rho$, $n_y=n_z=3$, with $\rho=1,2,4$
\item Courant factor $= 0.5$
\item Iterations: $100000\rho$, output every $100\rho$ iterations
  (corresponding to 1 crossing time)
\item Gauge: harmonic, i.e. 
  $\partial_t \alpha = -\alpha^2 \mbox{tr}K$, $\beta^i=0$
\end{itemize}

The initial data is given by 
\begin{eqnarray}
g_{ij} &=& \delta_{ij} + \varepsilon_{ij} , \\
K_{ij} &=& 0 , \\
\alpha &=& 1 , \\
\beta^i &=& 0 ,
\end{eqnarray}
where $\varepsilon_{ij}$ is a random number with a probability distribution,
which is uniform in the interval $(-10^{-10}/\rho^2,+10^{-10}/\rho^2)$. 
After a small number of timesteps, the random noise in $g_{ij}$ will have
propagated into all other evolved quantities, except for $\beta^i$, which
will remain identically zero for all time. Hence our initial data differ
slightly form the ones in \cite{Alcubierre2003:mexico-I},
who add random noise to all quantities which need initialization.

There is one obvious problem with the robust stability test. The
initial data are random, and we must carefully check that the random
number generator we use does not introduce errors because the numbers
are not totally random. We use C's ``random'' function on Linux Redhat 7.3,
a pseudo-random number generator based on a non-linear additive
feedback algorithm that avoids the short-comings of some of the older
implementations of the ``rand'' function.  
As a seed for the random number generator we use the time where the
subroutine is called plus the clock cycle of the processor on which it
is called. Since our code is parallelized and each processor uses its
own seed, the actual random numbers on the grid depend on the number
of processors. This could be avoided by additional coding, however,
the qualitative result of the robust stability test is not expected to
depend on the actual random numbers. We have run the test several
times on different numbers of processors. We have also tried increasing
the size of the domain, both in the x-direction and the y- and 
z-directions (which results in a different number of random numbers being
generated). We find that although this does change the results, many
features are robust and do not change with these variations. We will
only comment on these robust features of the test.

ADM and BSSN results are similar to those reported in
\cite{Alcubierre2003:mexico-I}. ADM grows exponentially 
(Fig.~\ref{fig_admrobust}), while BSSN is stable (Fig.~\ref{fig_bssnrobust}).
Note that ADM clearly shows the signature of a grid mode, since when
plotting versus the number of iterations almost identical exponential
growth is obtained for the three different resolutions.

In~\cite{Bona:2003qn}, the ADM system is run with a Courant factor 
of only $0.03$ with the result that it is stable for much longer, up
to 200 crossing times.
We have lowered the Courant factor in the ADM
run to $0.25$ and found that we do not see exponential growth in the
Hamiltonian constraint. When plotting $\mbox{tr}K$ we see some growth,
similar to Fig.~2 in~\cite{Bona:2003qn}, but we do not encounter a blowup
(we ran the coarsest resolution to 10000 crossing times, and did not
encounter exponential growth).  The results are shown in
Fig.~\ref{fig_admrobust025}. We also tried with Courant factor $0.03$
as in~\cite{Bona:2003qn}, and this did not show any exponential blowup (the
run was stopped at 600 crossing times). Preliminary experiments indicate
that the transition between stable and unstable Courant factors
for ADM lies between $0.4$ and $0.5$. This deserves further investigation.

The results for AA are shown in Fig.~\ref{fig_AArobust}. AA is
qualitatively stable for the 1000 crossing times suggested for this
test, but oscillations in the maximum and minimum of the Hamiltonian
occur. There are two types of oscillations, one with a long wavelength
and one with a short wavelength. We can see that the short wavelength
oscillations dampen out. The long wavelength of the medium and high
resolution runs seems to be the same, while the wavelength in the low
resolution run is about half of the wavelength in the other runs. As
pointed out earlier, these features are robust when changing the
domain-size, but the amplitude of the oscillations are affected by
this change.

Given the presence of these slow oscillations, we ran another set of tests
where we evolve for 10000 crossing times. 
Fig.~\ref{fig_AArobustlong} shows the results. We can see that at late
times, the constraint violation for all 3 resolutions are growing, and
this suggests that the AA system is stable for random noise initial
data for a long time, but it will eventually become unstable. We also
ran the test of the BSSN system to 10000 crossing times but found no
instabilities or any indication of longterm growing oscillations in
this case.

We tried lowering the Courant factor for the AA runs. We ran the
coarsest resolution to 10000 crossing times with a Courant factor of
0.25 and found that the growth in the Hamiltonian constraint still
happens, but it happens later than in the run with the Courant factor
of 0.5. Lowering the Courant factor for the AA system does not change
the general features of the oscillations.

\subsection{Gauge wave stability test}

In this test we look at the ability of the evolution systems to
handle gauge dynamics. This is done by considering flat Minkowski
space in a slicing where the 3-metric $g_{ij}$ is time dependent. The
gauge wave is then given by
\begin{eqnarray}
g_{xx} &=& 1 + A \sin \left( \frac{2 \pi (x - t)}{d} \right), \\
g_{yy} = g_{zz} &=& 1, \\
g_{xy} = g_{xz} = g_{yz} &=& 0, \\
K_{xx} &=& -\frac{\pi A}{d} \frac{ \cos \left( \frac{2 \pi (x - t)}{d}
\right) }{ \sqrt{1 + A \sin \left( \frac{2 \pi (x - t)}{d} \right) }
},\\
K_{ij} &=& 0 \qquad \textrm{ otherwise} \\
\alpha &=& \sqrt{1 + A \sin \left( \frac{2 \pi (x - t)}{d} \right)} , \\
\beta^i &=& 0.
\end{eqnarray}
Here $d$ is the size of the domain in the $x$-direction 
and $A$ is the amplitude of the wave.
Since this wave propagates along the $x$-axis and all derivatives are
zero in the $y$- and $z$-directions, the problem is essentially one
dimensional. 

These are the parameters we use in our gauge stability tests:
\begin{itemize}
\item Simulation domain: $x \in [-0.5, +0.5]$
\item Points: $n_x=50\rho$, $n_y=n_z=3$, $\rho=1,2,4$
\item Courant factor $= 0.25$
\item Iterations: $200000\rho$, output every $200\rho$ iterations
  (corresponding to 1 crossing time)
\item $A=0.1$ and $A=0.01$ 
\item Gauge: harmonic, i.e. 
  $\partial_t \alpha = -\alpha^2 \mbox{tr}K$, $\beta^i=0$
\end{itemize}

Since we know the analytical solution at all times, we can compare our
numerical results to the analytical results, see
also~\cite{Calabrese02a}. As well as testing if the system has
exponentially growing modes, we can check the convergence
of the numerical result to the analytical solution with increasing
resolution.

Fig.~\ref{fig_AAgauge} shows that for a gauge wave amplitude
$A=0.1$ the AA system converges as expected for a finite time
interval, but there is exponential growth and the run crashes after about
100 crossing times. 
For $A=0.01$ (Fig.~\ref{fig_AAgaugeA001}) it takes longer
for this non-convergence to show, about 1000 crossing times, 
but it is still there. If the AA
system together with the given gauge equations 
was symmetric hyperbolic, and if one had a stable
discretization scheme, then the result would be
convergent. Conversely, assuming that ICN is an appropriate
discretization scheme, we would conclude that the AA system 
in harmonic gauge is not symmetric hyperbolic, which agrees with our 
result in Sec.~\ref{sec:hyper}.

The ADM system shows no growth of the constraints at all, and the
constraint violation remains at machine precision. This is probably
because for a 1-dimensional gauge wave the ADM system simplifies such
that no constraints violating modes are possible.

The results for the BSSN system are shown in
Fig.~\ref{fig_bssngauge}. The BSSN system becomes unstable and
non-convergent rather quickly (on the order of 100 crossing
times). This is somewhat surprising since the BSSN system has been
very successful in single black hole evolutions and neutron star
evolutions~\cite{Alcubierre02a,Duez:2002bn}, so we expect it to be
able to handle gauge dynamics. However, part of the robustness of BSSN
can be attributed to the fact that constraint violating modes have a
finite speed~\cite{Alcubierre99e,Knapp02a} and can propagate out of
the grid for example for radiative boundary conditions. However, on
our grid with periodic boundaries, when a constraint violating mode
appears it will stay on the grid, which is probably the reason that
the BSSN system fails this test. Thus, this test by itself cannot
determine whether a system can handle gauge dynamics, but must be
accompanied by other tests without periodic boundaries. In particular,
there is no contradiction to the observed stability of BSSN for single
black holes with radiative boundaries.

\subsection{Linear wave stability test}

In this section we investigate the ability of the evolution systems to
propagate the amplitude and phase of a traveling gravitational
wave. The amplitude of this wave is sufficiently small so that we are
in the linear regime. This test reveals effects of numerical
dissipation and other sources of inaccuracy in the evolution
algorithm.

The initial 3-metric and extrinsic curvature are obtained from
\begin{equation}
  ds^2= - dt^2 + dx^2 + (1+b) \, dy^2 + (1-b)  \, dz^2,
\end{equation}
where 
\begin{equation}
   b  =  A \sin \left( \frac{2 \pi  (x-t)}{d}\right),
\end{equation}
$d$ is the size of the propagation domain, and $A$ is the amplitude of
the wave. The extrinsic curvature tensor is given by
\begin{eqnarray}
   K_{yy} &=& -\frac{A \pi}{d} \cos\left(\frac{2 \pi x}{d}\right) , \\
   K_{zz} &=& \frac{A \pi}{d} \cos\left(\frac{2 \pi x}{d}\right) , \\
   K_{ij} &=& 0, \qquad \text{otherwise}.
\end{eqnarray}

These are the parameters of our run:
\begin{itemize}
\item Simulation domain: $x \in [-0.5, +0.5]$
\item Points: $n_x=50\rho$, $n_y=n_z=3$, $\rho=1,2,4$
\item Courant factor $= 0.25$
\item Iterations: $200000\rho$, output every $200\rho$ iterations
  (corresponding to 1 crossing time)
\item $A=10^{-8}$ 
\item Gauge: geodesic, i.e.\ $\alpha = 1, \beta^i = 0$.
\end{itemize}

Figs.~\ref{fig_linearcomp},\ref{fig_linearnorm}, show the results for
the ADM, AA, and BSSN systems, respectively.
Fig.~\ref{fig_linearcomp} is a comparison of the numerical wave to the
analytical wave at 100 crossing times. We see that the AA numerical
wave travels slightly faster than the ADM and BSSN numerical
waves. Fig.~\ref{fig_linearnorm} shows the
$L_2$-norm of the difference between the numerical and analytical
linear waves as a function of time, and again the wave in the AA
system travels faster than in the other systems. The
Hamiltonian constraint has a value of about $10^{-8}$ at the end of
the run for the AA system, so we are still well within the linear
regime. It is also worth noting that in the ADM system the constraint
violation is constant throughout the evolution, while the constraints
grow slightly for the other two systems (but the maximum violation is
of the order $10^{-8}$ for the entire run, for all resolutions).

\section{Numerical tests involving a single Schwarzschild black hole}
\label{secBH}

In this Section we present numerical results for the evolution of a single
Schwarzschild black hole in two different gauges.

\subsection{Geodesic slicing}

The initial data for this test is a Schwarzschild black hole in
isotropic coordinates. We use geodesic slicing to evolve the initial
data. In geodesic slicing, the coordinate lines correspond to freely
falling observers, and the resulting slicing of Schwarzschild can be
expressed analytically in terms of Novikov coordinates,
e.g.~\cite{Frolov98}, Chapter 2.7.2, and~\cite{Bruegmann96}, which we
transform to isotropic coordinates for direct comparison with the
numerical results. To this end, we numerically solve the following
implicit equation for the Schwarzschild area radial
coordinate $R=R(\tau,R_\text{max})$,
\begin{eqnarray}
\label{Roftau_Rmax}
\tau &=& \frac{R_\text{max}}{2 M} \left( \sqrt{\frac{R}{2 M} \left(1 -
  \frac{R}{R_\text{max}}\right)} \right.+ \nonumber \\
&\quad& \quad \qquad \left. \sqrt{\frac{R_\text{max}}{ 2 M }}
  \arccos{\sqrt{\frac{R}{R_\text{max}}}} \right), 
\end{eqnarray}
where $R = R_\text{max}$ is the position at $\tau = 0$ of a freely
falling observer starting at rest, and $M$ is the mass of the black hole. 
To transform from the $R_\text{max}$ radial coordinate to the
isotropic radial coordinate $r$ we use
\begin{equation}
R_{\text{max}}({r}) = \frac{\left(M + 2 {r}\right)^2}{4 {r}} .
\end{equation}
Then the ${r}{r}$-component of the metric in isotropic coordinates 
is given by
\begin{equation}
\label{eq:novikovg}
g_{{r}{r}} = \psi({r})^4 
\left(\frac{\partial R}{\partial R_\text{max}} \right)^2 ,
\end{equation}
where
\begin{eqnarray}
\frac{\partial R}{\partial R_\text{max}} &=& \frac{3}{2} - \frac{1}{2}
\frac{R}{R_{\text{max}}} \nonumber + \\ 
&\quad& \frac{3}{2} \sqrt{ \frac{R_{\text{max}}}{R}
  - 1}\; \arccos{\left(\sqrt{\frac{R}{R_{\text{max}}}}\;\right)} 
\end{eqnarray}
and $R$ is computed from Eq.~(\ref{Roftau_Rmax}).
Here $\psi({r}) = 1 + \frac{M}{2 {r}}$. 
Note that there is a typo in \cite{Bruegmann96},
Eq.~(16).

Analytically, the metric becomes infinite at time
\begin{equation}
t_\text{crash} = \pi M,
\end{equation}
which is the time for an observer that starts from rest at the horizon
to reach the Schwarzschild singularity.
This results in a crash in the numerical computations.  We run this
test on a 3D grid in the so-called cartoon mode
\cite{Alcubierre99a}, because the spherical symmetry of the problem
means that we can do a computation using information only on a
line passing trough the center of the black hole.  The computational
domain is $z \in [0, 80M ], x=0, y=0$.  We use second order accurate
finite differencing together with an iterative Crank-Nicholson scheme
with a Courant factor of $0.25$ for evolution.  Since we never run
longer than $t_\text{crash} = \pi M$, the outer boundary at $80M$ has
no effect on the black hole located at the origin.

In order to compute the order of convergence, 
\begin{equation}
\sigma = \log_2 \left| \frac{f_{4 h} - f_{2 h}}{f_{2 h} - f_h} \right| ,
\end{equation}
we run the test at the resolutions $h=0.01M$, $2h=0.02M$, and
$4h=0.04M$.  We compute $\sigma$ on each grid-point present in the
coarsest run, and $f_{h}$ is the value of the quantity under
consideration (here the metric component $g_{zz}$) for resolution $h$.
Our grid-points are not in the same places for all three resolutions,
because we always stagger the puncture between two grid-points. This
means that we must interpolate to get the values at all the coarse
grid-points, for which we use 4th order Lagrange interpolation.
Fig.~\ref{fig_sigma} shows $\sigma$ for the AA system
at time $t=\frac{\pi}{2}M$ and
time $t=3.0M$ in the region close to the black hole where the metric
deviates the most from the flat conformal metric. We see that we have
second order convergence close to the black hole in both cases. The
spikes in the curve at later times are due to the fact that the curves
of $g_{zz}$ for the 3 resolutions cross.

\subsection{1+log lapse, gamma driver shift}

We have implemented 1+log lapse and gamma driver shift in our
code. This gauge choice makes the BSSN system stable for more than
$1000M$ for certain single black hole runs~\cite{Alcubierre02a}. While
the geometric motivation of this gauge choice, namely singularity
avoidance and reduction of slice-stretching, should be valid for any
evolution system, it is unclear whether the AA system will be as
stable as BSSN or unstable in this test case. In particular, note that
the gamma driver shift is designed to control the evolution of one of
the BSSN variables, $\tilde\Gamma^i$. 
We have not implemented proper outer boundary conditions for the AA
system, so in our simulations we have waves coming in from the outer
boundaries. We use the same parameters as for the geodesic runs, i.e.\
the domain in cartoon mode is $z \in [0, 80M ], x=0, y=0$,
the resolution is $ h = 0.01M$, and we use a Courant factor of $0.25$.

Our AA run crashes at around $20 M$. Very steep gradients appear near the
black hole and they eventually kill the run. The reason for this seems
to be that the 1+log lapse depends on the trace of the physical
extrinsic curvature. Since we evolve the conformal extrinsic curvature,
finite difference errors are enlarged by a factor of $\psi^4$ when
computing the physical extrinsic curvature, and these account for the
differences in the lapse evolution between BSSN and AA runs. When
using the AA system, the lapse drops very fast, leaving a sharp
gradient between the frozen region and the region where the fields can
evolve. The gradients in the physical variables that kill the run
appear where the frozen region meets the dynamic region. In the BSSN
system this gradient is more shallow and we do not see this effect.

This demonstrates that, as expected, a gauge choice that leads to
numerically stable evolutions with one evolution system may well be
unstable for others.

\section{Discussion}
\label{secDiscuss}
\label{Discussion}

We have investigated the numerical properties of the previously
untested AA evolution system by a variety of numerical experiments,
which for comparison we also performed for the ADM and BSSN systems.
Since one aspect of the test suite put forward in
\cite{Alcubierre2003:mexico-I} is that different numerical
implementations of one and the same evolution system can be compared,
let us point out that our results for ADM and BSSN agree with
\cite{Alcubierre2003:mexico-I} for the specific results shown there (see also
\cite{MayaBSSN_web}). 

\begin{figure*}[ht]
\begin{center}
\epsfig{file=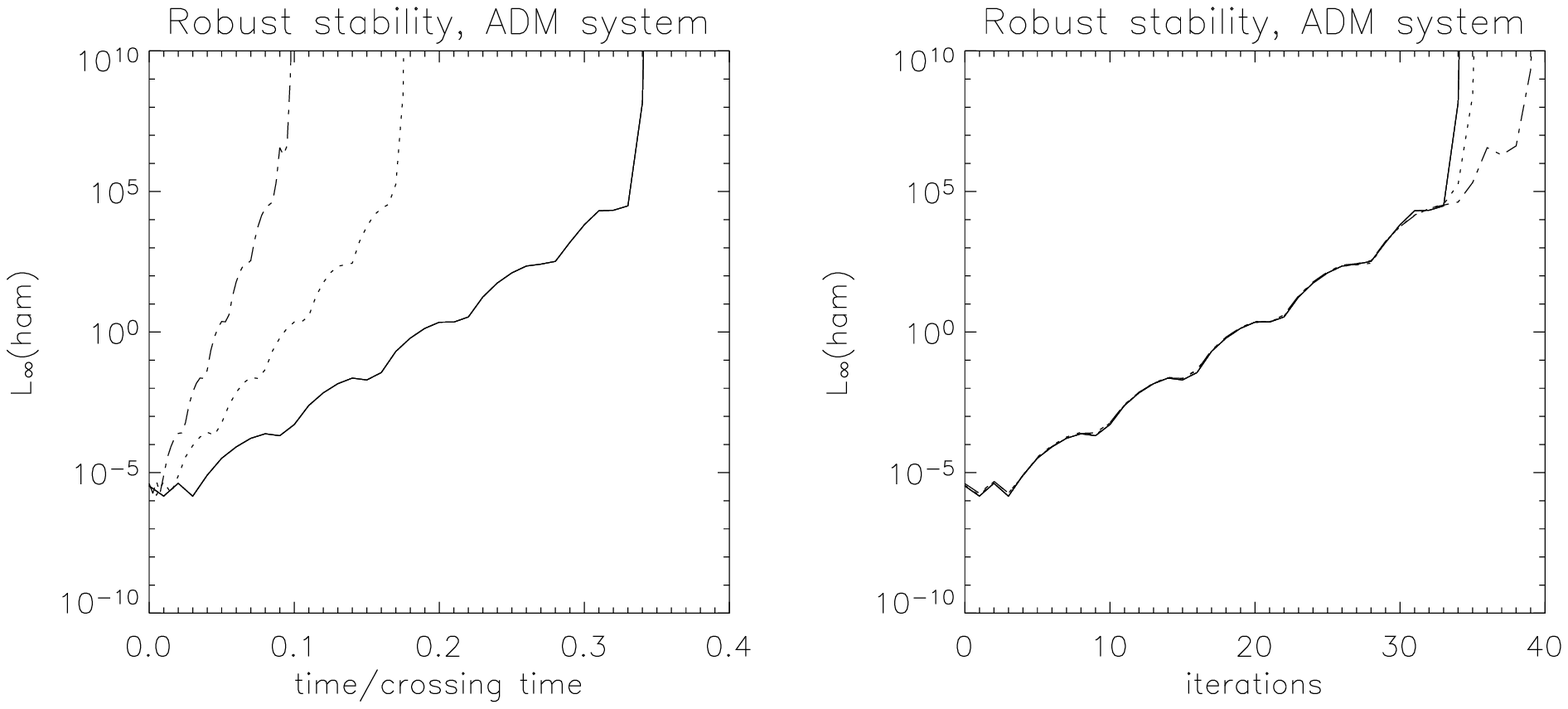, width=15cm}
\end{center}
\caption[]{\label{fig_admrobust}  Robust stability test
  for the ADM system. The legend is: solid: $\rho=1$, dotted:
  $\rho=2$, dash-dot: $\rho=4$, $L_\infty(\text{ham})$ is shown both
  as a function of time (left) and iterations (right). Note the
  almost perfect alignment when plotting versus the number iterations,
  which is the signature of a grid mode.}
\end{figure*}

\begin{figure*}[ht]
\begin{center}
\epsfig{file=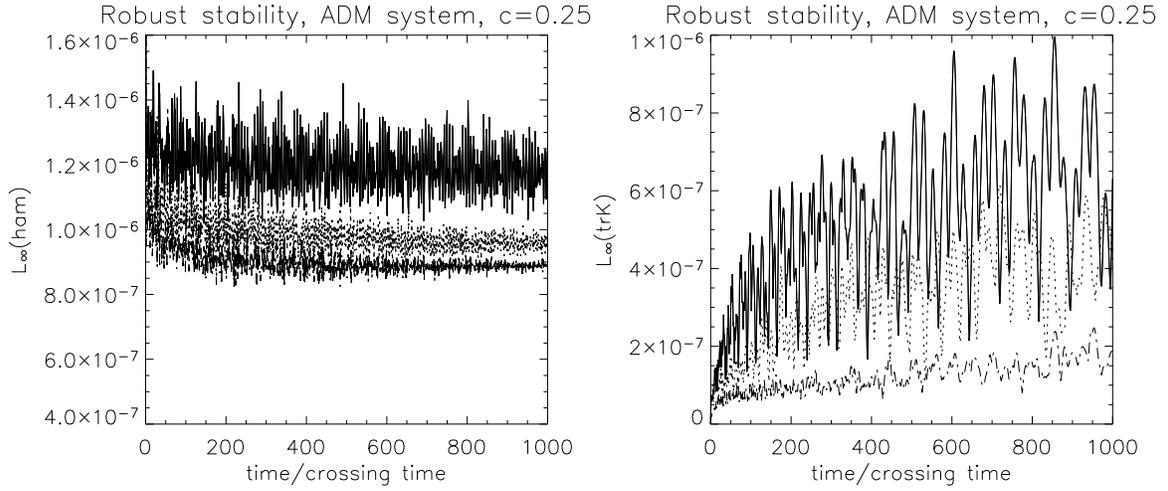, width=15cm}
\end{center}
\caption[]{\label{fig_admrobust025} Robust stability test for the ADM
  system, with Courant factor $0.25$. The legend is: solid: $\rho=1$, dotted:
  $\rho=2$, dash-dot: $\rho=4$. We show  $L_\infty(\text{ham})$ on the
  left and  $L_\infty(\text{tr$K$})$ on the right. The exponentially
  growing mode seen in Fig.~\ref{fig_admrobust} is not present.}
\end{figure*}

\begin{figure*}[ht]
\begin{center}
\epsfig{file=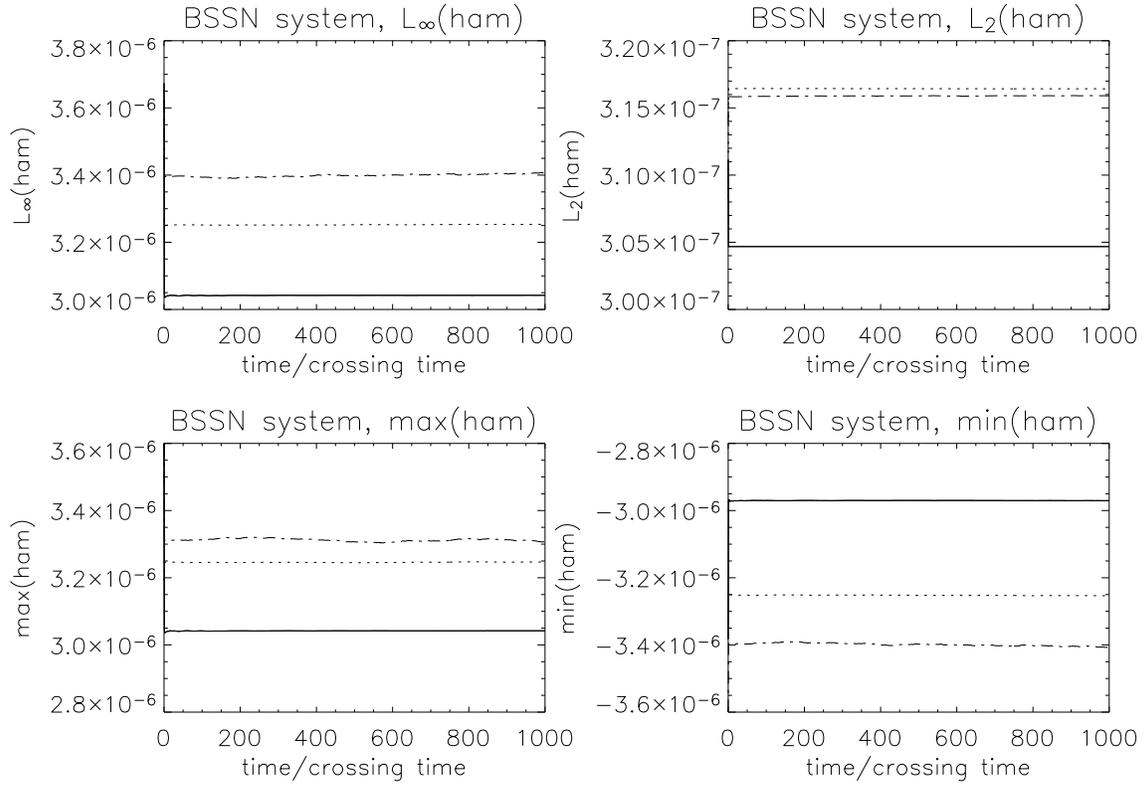, width=15cm}
\end{center}
\caption[]{\label{fig_bssnrobust} Robust stability test
  for the BSSN system. The legend is: solid: $\rho=1$, dotted:
  $\rho=2$, dash-dot: $\rho=4$.}
\end{figure*}

\begin{figure*}[ht]
\begin{center}
\epsfig{file=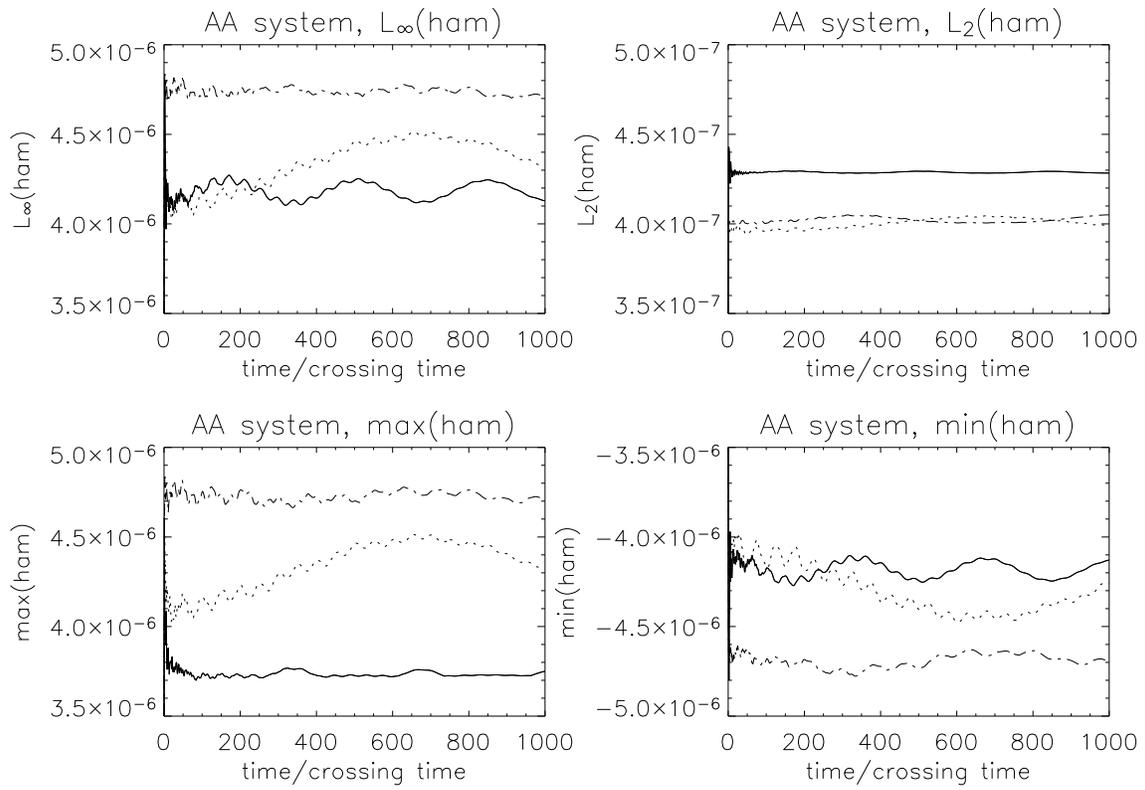, width=15cm}
\end{center}
\caption[]{\label{fig_AArobust} Robust stability for the
  AA system, run until 1000 crossing times.
  The legend is: solid: $\rho=1$, dotted:
  $\rho=2$, dash-dot: $\rho=4$.}
\end{figure*}

\begin{figure*}[ht]
\begin{center}
\epsfig{file=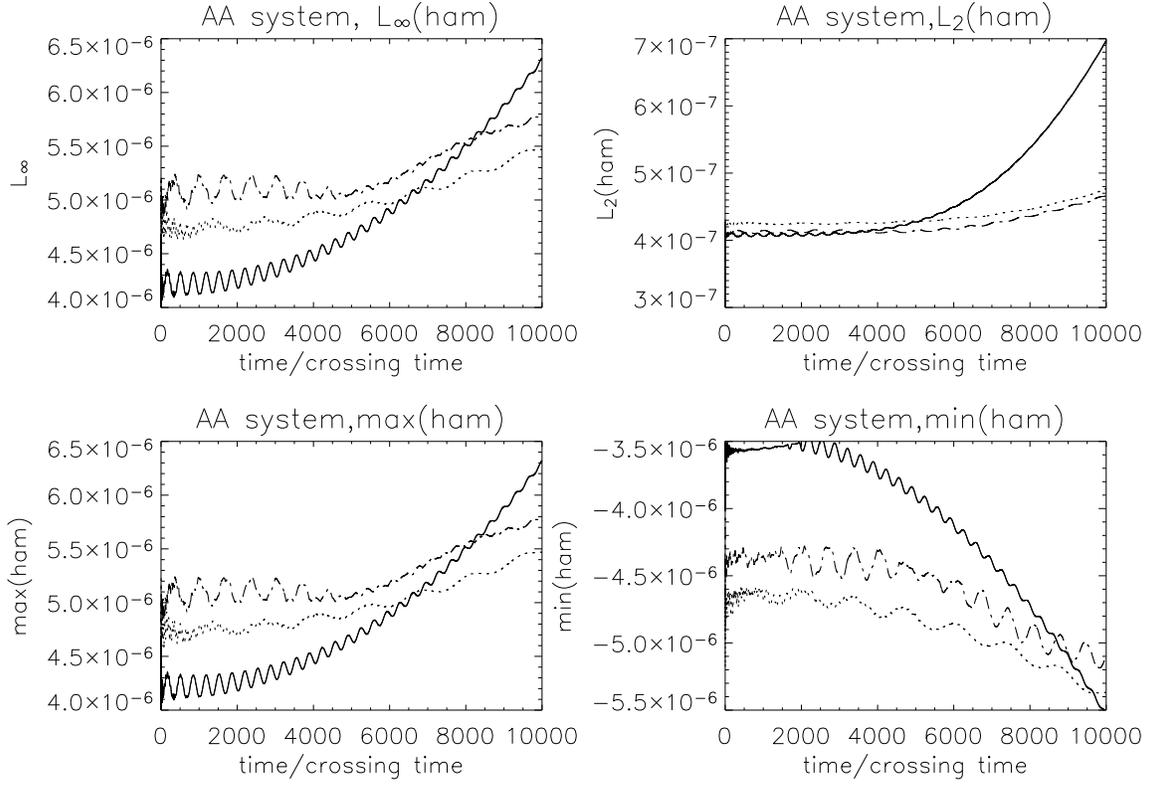, width=15cm}
\end{center}
\caption[]{\label{fig_AArobustlong} Robust stability for
  the AA system, run until 10000 crossing times. The legend is:
  solid: $\rho=1$, dotted: $\rho=2$, dash-dot: $\rho=4$. We see that at
  late times the constraint violations are growing, which will
  probably cause a crash if we evolve long enough.}
\end{figure*}

\begin{figure*}[ht]
\begin{center}
\epsfig{file=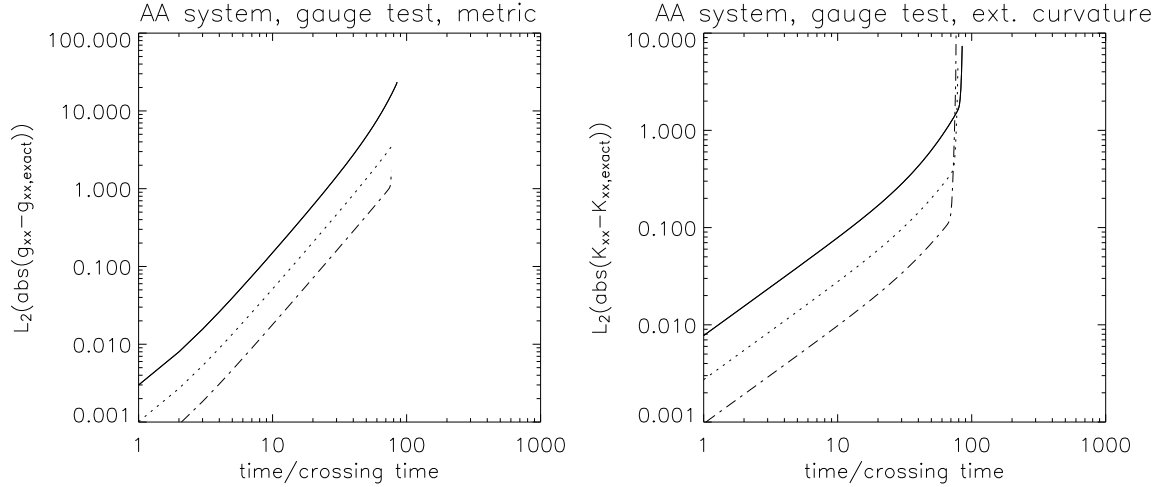, width=15cm}
\end{center}
\caption[]{\label{fig_AAgauge} Gauge stability test for
  the AA system with a wave amplitude $A=0.1$. The legend is:
  solid: $\rho=1$, dotted: $\rho=2$, dash-dot: $\rho=4$. As expected,
  the metric converges at early times, but there is exponential growth
  and the run crashes after about 100 crossing times.}
\end{figure*}

\begin{figure*}[ht]
\begin{center}
\epsfig{file=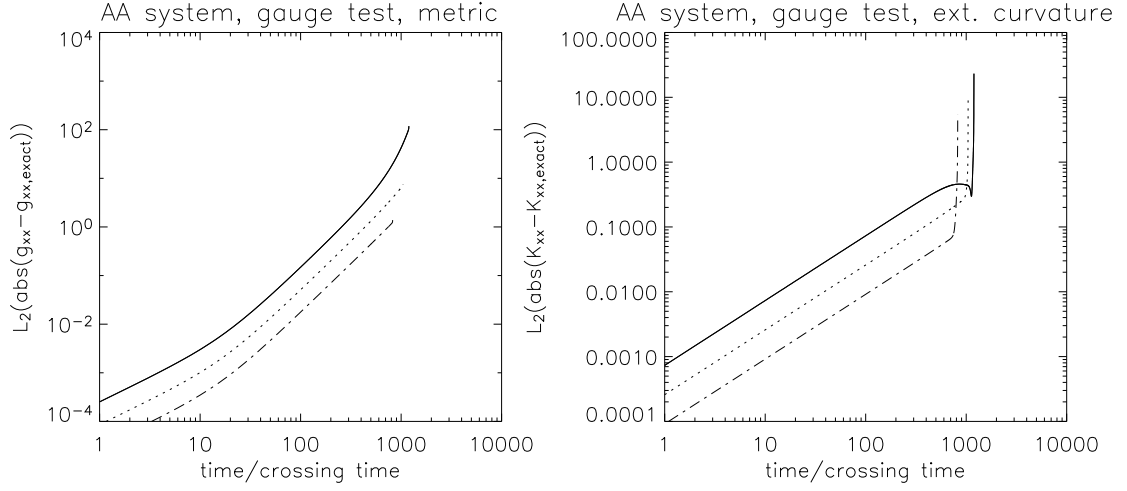, width=15cm}
\end{center}
\caption[]{\label{fig_AAgaugeA001} Gauge stability test
  for the AA system with a wave amplitude $A=0.01$. The legend is:
  solid: $\rho=1$, dotted: $\rho=2$, dash-dot: $\rho=4$. 
  We see that with this smaller amplitude the runs last 10 times longer but 
  still crash eventually.}
\end{figure*}

\begin{figure*}[ht]
\begin{center}
\epsfig{file=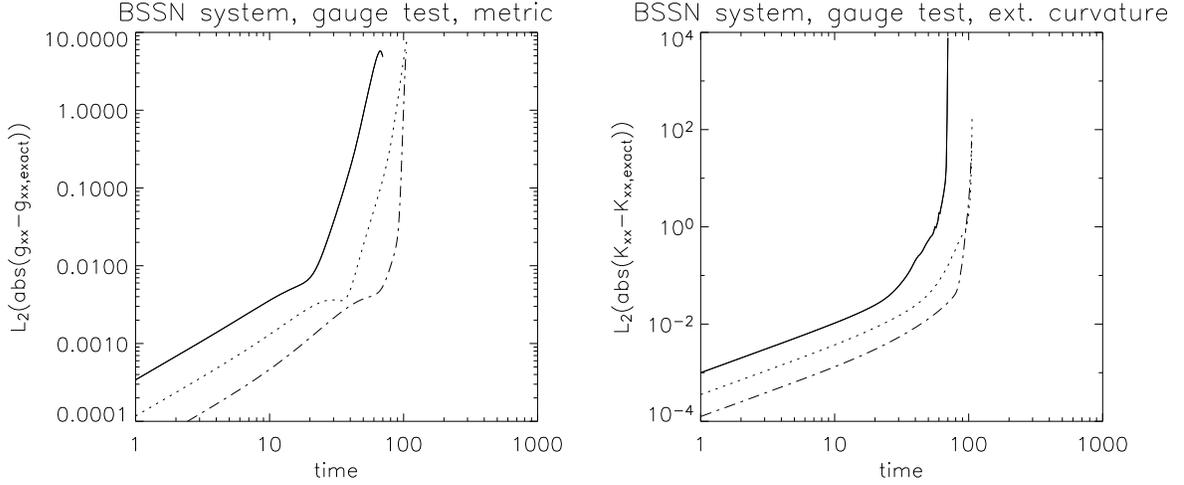, width=15cm}
\end{center}
\caption[]{\label{fig_bssngauge} Gauge stability test
  for the BSSN system with a wave amplitude $A=0.01$. The legend is:
  solid: $\rho=1$, dotted: $\rho=2$, dash-dot: $\rho=4$. Both the
  metric and the extrinsic curvature tensor become non-convergent in a
  short time.}
\end{figure*}

\begin{figure*}[ht]
\begin{center}
\epsfig{file=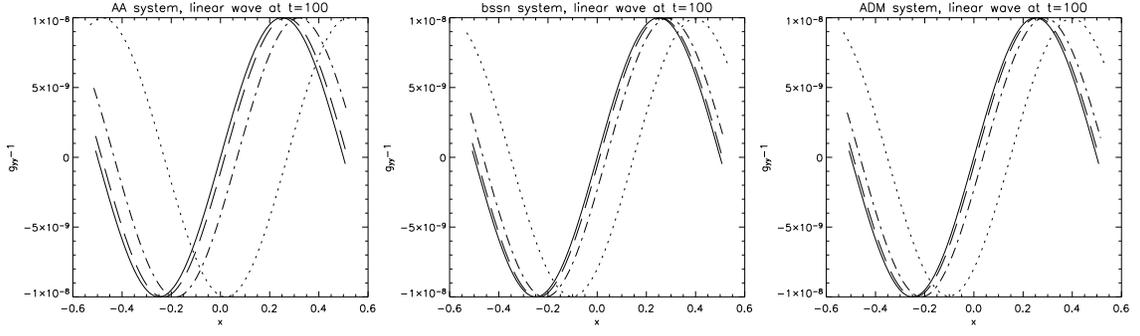, width=15cm}
\end{center}
\caption[]{\label{fig_linearcomp} Linear wave stability test for (from
  left to right) the AA system, the BSSN system, and the ADM system. We
  compare the numerical wave at 100 crossing
  times to the analytical solution. The legend is: solid: analytical
  solution, dotted: $\rho=1$, dash-dot: $\rho=2$, dash: $\rho=4$.}
\end{figure*}

\begin{figure*}[ht]
\begin{center}
\epsfig{file=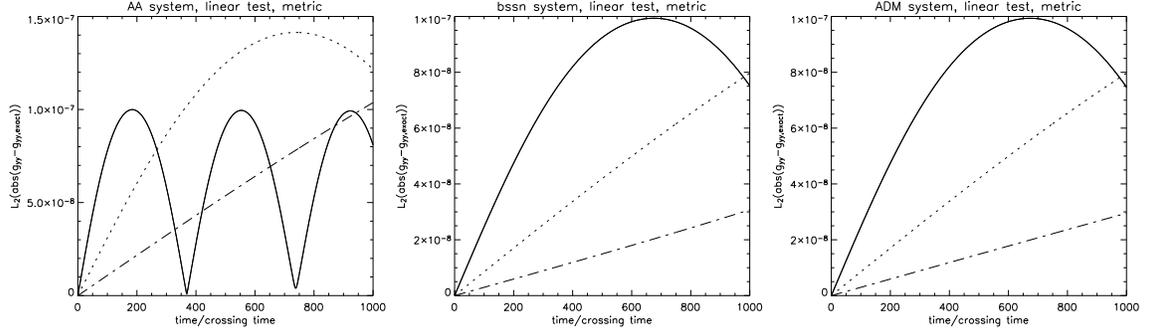, width=15cm}
\end{center}
\caption[]{\label{fig_linearnorm} Linear wave stability test for (from
  left to right) the AA system, the BSSN system, and the ADM system. We
  show the $L_2$ norm of the difference between the analytical and
  numerical values at different crossing times. The legend is: solid:
  $\rho=1$, dotted: $\rho=2$, dash-dot: $\rho=4$.}
\end{figure*}

\begin{figure*}[ht]
\begin{center}
\epsfig{file=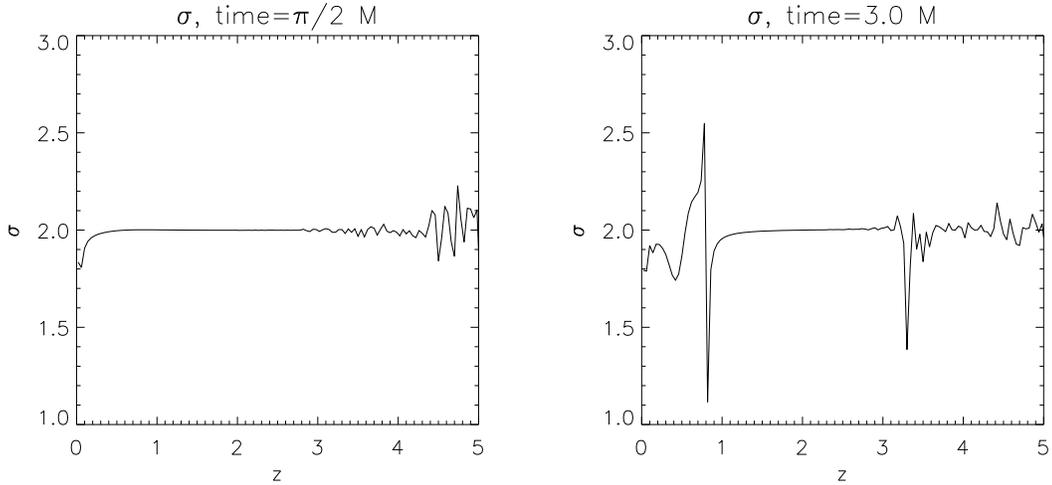,width=15cm}
\end{center}
\caption[]{\label{fig_sigma}
  Geodesic slicing of Schwarzschild.  Shown is the order of
  convergence $\sigma$ of the conformal metric component $\bar g_{zz}$
  for the AA system in the computational domain near the black hole
  for $t=\frac{\pi}{2}M$ and $t= 3.0M$. We observe second order
  convergence near the black hole.  }
\end{figure*}

However, if the Courant factor in the robust stability test is lowered
from $0.5$ (as suggested in~\cite{Alcubierre2003:mexico-I}) to $0.25$
we find that the stability of ADM dramatically improves (compare
Figs.~\ref{fig_admrobust} and \ref{fig_admrobust025}), while the
stability properties of AA remain unchanged.  This implies
that the Courant factor used in \cite{Alcubierre2003:mexico-I} is too
large for the ADM system causing immediate exponential growth, instead of the
initially linear growth expected for a weakly hyperbolic system. This
confirms a similar observation already described
in~\cite{Bona:2003qn}. We use iterative Crank-Nicholson in all
our simulations, but~\cite{Bona:2003qn} also point out that
dissipation can mask the linear growth expected for
ADM. So one should repeat these experiments with a less dissipative
scheme like third order Runge-Kutta to look for linear growth in ADM,
but also in the AA systems, see Sec.~\ref{sec:hyper}.

One property of BSSN that has not been explicitly stated
in~\cite{Alcubierre2003:mexico-I} is its drastic failure on periodic
domains. This observation is consistent with, and actually
strengthens, the notion that BSSN performs well because it is able to
propagate modes off the grid~\cite{Alcubierre99e}, in particular for
isolated systems with radiative boundary conditions. It will be
interesting to see how AA, ADM, and BSSN perform on a gauge wave test
with outer boundary conditions that let the gauge wave propagate away
from the domain.

Our findings for the AA system can be summarized as follows.
In the robust stability test, the AA system is stable for a long time,
but eventually does go unstable. In this case the AA system does not do
as well as the BSSN system. For gauge waves in periodic geometry, the
AA system is unstable, but the runs last much longer than the
corresponding BSSN runs. The linear wave test
shows that the AA system produces a larger drift in the phase than the
ADM and BSSN systems, causing the linear wave to propagate faster
compared to the analytical solution in the AA evolution than in the
other two evolutions.

In the black hole runs, AA does as well as ADM and BSSN for
geodesic slicing runs, but fails when trying to use gauges that makes
BSSN runs long term stable. There are two distinct aspects of this
gauge choice. While the gauge is appropriate geometrically
(singularity avoiding and countering slice-stretching), there is no reason
to expect numerical stability for the AA system. Since it was
non-trivial to find a numerically stable gauge choice for BSSN, it is
not too surprising that additional work is required to find a gauge
choice for the AA system that allows long run times for a single black
hole, if indeed such a choice exists.

We have found that the tests published in
\cite{Alcubierre2003:mexico-I} are helpful in exploring the
properties of the AA system, but also that, not unexpectedly, these
tests cannot by themselves determine whether an evolution system is
worth exploring further for a particular physical system, say black
hole evolutions. Note that one should expect that for different
physical initial data sets different evolution systems are optimal,
see e.g.~\cite{Lindblom:2002et}.  Since the relationship between
choice of evolution system, gauge choice, outer boundary conditions,
and the physical properties of the problem we are trying to simulate
is complicated, it is not clear that sufficient understanding of the
numerics needed to evolve binary black holes can be gained by singling
out for example the evolution system and ignoring the other issues
involved. Ultimately, if one wants to evolve black holes, one should
evolve black holes.

While the AA system has had some success in our numerical
experiments, tests with black holes, proper outer boundaries, and
appropriate gauge choices are needed to determine if the AA system has
a future in numerical relativity.

\begin{acknowledgments}
We would like to thank A. Alekseenko and D. Arnold for helpful
discussions. 
We acknowledge the support of the Center for Gravitational Wave Physics
funded by the National Science Foundation under Cooperative Agreement
PHY-01-14375. This work was also supported by NSF grants PHY-02-18750
and PHY-98-00973.
\end{acknowledgments}

\bibliography{references,AA}

\newpage


\begin{figure*}[ht]
\begin{center}
\epsfig{file=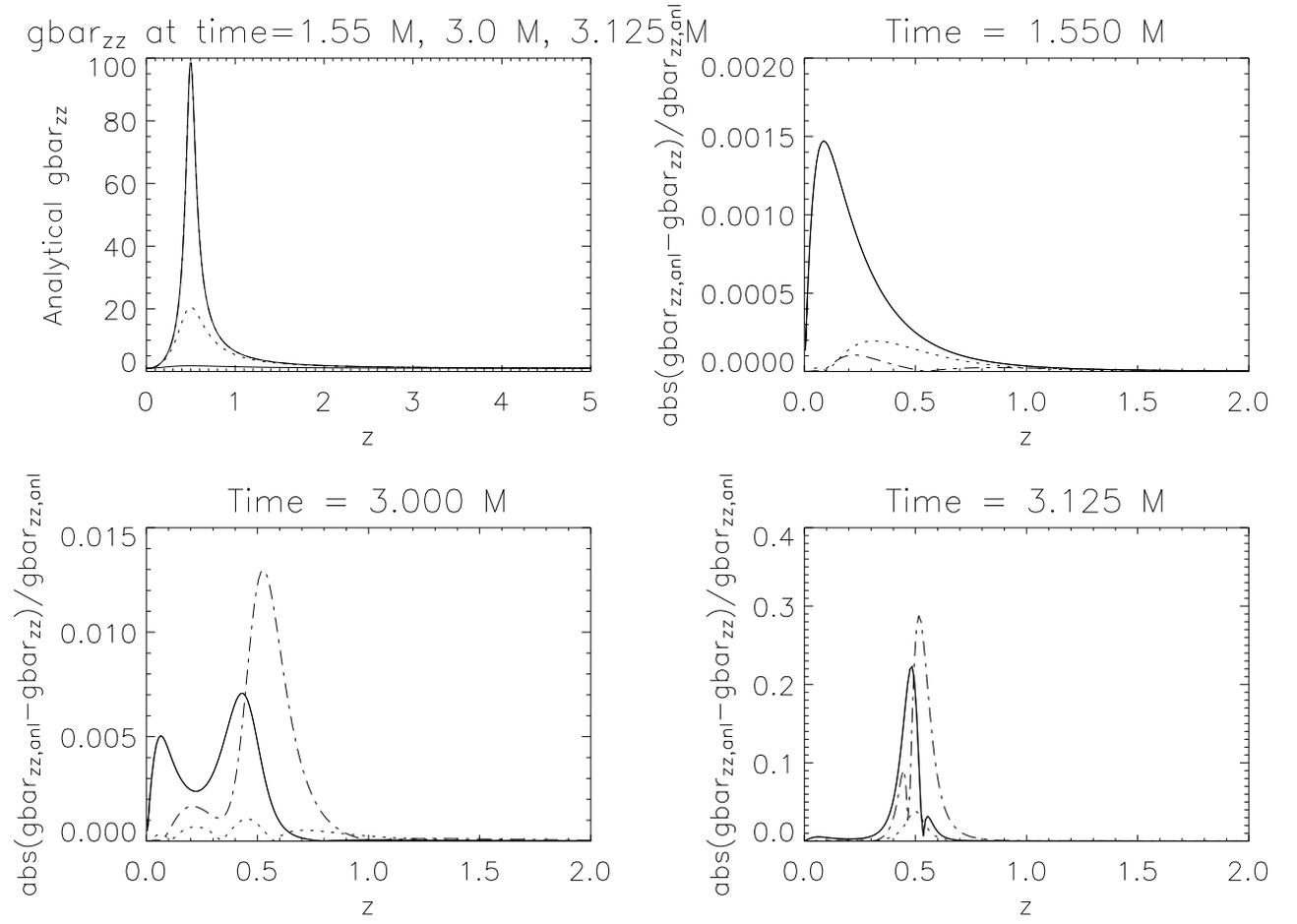}
\end{center}
\caption[]{\label{fig_crash} 
Geodesic slicing of Schwarzschild. The first panel shows the
analytical conformal metric component $\bar g_{zz}$ as a
function of $z$ in the inner region of the domain (same region as in
Fig.~\ref{fig_sigma}). The analytical solution is shown at 3 different
times: $t=1.55$ (solid), $t=3.0$ (dotted) and $t=3.125$
(dash-dot). The other 3 panels show the absolute value of the relative
difference of the numerical and analytical solutions at the 3
different times. In these plots the AA solution is solid, ADM is
dotted, and BSSN is dash-dot.}
\end{figure*}

\end{document}